\documentclass[twocolumn]{svjour3}
\usepackage{shadethm}
\spnewtheorem{observation}{Observation}{\it}{\rm}

\def\M{\hbox{$\mathcal{M}$}}
\def\A{\hbox{$\mathcal{A}$}}
\def\P{\hbox{$\mathcal{P}$}}
\def\E{\hbox{$\mathcal{E}$}}
\def\K{\hbox{$\mathcal{K}$}}

\def\PREC{\prec_{\mathcal{P}}}
\usepackage{epigraph}
\usepackage[tight,footnotesize]{subfigure}
\usepackage{pbox}
\usepackage{latexsym}
\usepackage{amsmath}
\usepackage{amssymb}
\usepackage{graphicx}
\usepackage[pdftex,
            pagebackref=true,
            colorlinks=true,
            linkcolor=blue
           ]{hyperref}
\usepackage{float}
\usepackage{alltt}
\usepackage{authblk}

\begin{document}

\title{On environments as systemic exoskeletons: \\Crosscutting optimizers and antifragility enablers}
\author{Vincenzo De Florio}
%\affil{MOSAIC research group\\
\institute{MOSAIC research group\\
       University of Antwerp \& iMinds research institute\\
       Middelheimlaan 1, 2020 Antwerp, Belgium\\
       \email{vincenzo.deflorio@gmail.com}\\
       \email{vincenzo.deflorio@uantwerpen.be}} \date{} \maketitle % 

\begin{abstract}
Classic approaches to General Systems Theory often adopt an individual perspective and a limited 
number of systemic classes. As a result, those classes include a wide number and variety of systems that 
result equivalent to each other. This paper presents a different approach: First, systems belonging to a same 
class are further differentiated according to five major general characteristics. This introduces a ``horizontal 
dimension'' to system classification. A second component of our approach considers systems as nested 
compositional hierarchies of other sub-systems. The resulting ``vertical dimension'' further specializes the 
systemic classes and makes it easier to assess similarities and differences regarding properties such as 
resilience, performance, and quality-of-experience.
Our approach is exemplified by considering a telemonitoring system designed in the framework of Flemish project
``Little Sister''. 
We show how our approach makes it possible to design intelligent environments able to closely follow a system's 
horizontal and vertical organization and to artificially augment its features by serving as crosscutting 
optimizers  and as enablers of antifragile behaviors.
\end{abstract}

%Key words: General Systems Theory. Systems Classification. Cybernetics. Intelligent Environments.

\section{Introduction}\label{s:intro}
%Two are the contributions of this paper.
%
%First, we propose a general systems classification based on two ``dimensions''...

Classic approaches to General Systems Theory (GST) such as the ones introduced in~\cite{RWB43} and~\cite{Bou56}
only consider a single, ``horizontal'' dimension. Boulding, for instance, classifies
systems through ``flat'' systemic classes: a system may be regarded as a ``Thermostat'', or
a ``Cell'', or a ``Plant'', and so forth, though all systems belonging to any given class
are no further differentiated. 
%We propose a simple differentiation method based on 
%%five major general characteristics of systems.

A second aspect shared by classical general systems classifications is the individual
and atomic perspective. In all behavioral classes introduced in~\cite{RWB43} and all
but one of those defined, for example, in~\cite{Bou56} systems are
considered as atomic, non-dividable elements.
The only exception to this rule is Boulding's class of social organizations, which is defined
as ``a set of roles  tied together with channels of communication'', though it is no further
analyzed.

A first contribution of this paper is the introduction
of a novel approach to general systems classification.
Following our approach, systems belonging to a same class are
differentiated according to five major general characteristics.
This introduces a ``horizontal dimension'' to system classification.
A second component of our approach is introduced through the assumption
that all systems should be considered as organizations of collective systems.
Such a recursive definition translates into a \emph{nested compositional hierarchy of sub-systems}, namely
``a pattern of relationship among entities based on the \emph{principle of increasing inclusiveness},
so that entities at one level are composed of parts at lower levels and are themselves nested
within more extensive entities''~\cite{HT:TE14a}. From said assumption we derive
the second,
``vertical'' classification dimension of our approach:
at the same time, systems are considered as either systems-of-systems or network-of-networks,
namely networks of nodes each of which may be another network.
Each of those nodes is a system, eligible thus to be classified along our horizontal
and vertical dimensions. Our stance is that a fair comparison of any two systems,
say $a$ and $b$, with respect to their features and emerging properties, should
be done by considering those two dimensions, up to some agreed upon level of detail or scale.

A discussion of our classification is
given in Sect.~\ref{s:clas} while in Sect.~\ref{s:res} we briefly consider how
our classification may be used in comparing the resilience of two systems.

%%%%%%%%%%%%%%%%%%%%%%%%%%%%%

The horizontal and vertical dimensions of our classification system are also one of the key
characteristics of a distributed hierarchical organization called Fractal Social Organization (FSO).
As in our general systems model, also FSO's~\cite{DF13c,DeFPa15a,DF15a} are a nested compositional
hierarchy of nodes. Such nodes are building blocks of a complex organization and are
called Service-oriented Communities~\cite{DeBl10,DFCBD12}.

Section~\ref{s:fso} briefly recalls the major elements
of FSO's.
A second contribution of this paper is the idea to make use of
the FSO organization
to design servicing infrastructures mimicking a system's vertical dimension and interfacing
its ``horizontal'' components~\cite{DFSB13a,DF15c}.
This may be used, e.g., to create intelligent environments
able to empower a community subjected to a natural or human-induced disaster~\cite{DFSB14}.
%In Sect.~\ref{s:case} we discuss a simple case study that we intend to model and analyze
%through a multi-agent simulation systems.

Our conclusions and a view to future work are finally stated in Sect.~\ref{s:end}.

%%%%%%%%%%%%%%%%%%%%%%%%%%%%%%%%%%%%%%%%%%%%%%%%%%%%%%%%%%%%%%%
\section{Two dimensions of system classification}\label{s:clas}
%%%%%%%%%%%%%%%%%%%%%%%%%%%%%%%%%%%%%%%%%%%%%%%%%%%%%%%%%%%%%%%
In Sect.~\ref{s:intro} we observed how traditional GST's mostly define
``flat'' classes of systems, and that said systems are often considered as
individual, atomic (i.e., non-decomposable) systems.
A reason for this is possibly that traditional theories are based on one or more
\emph{systemic touchstones}, which we defined in~\cite{DF14c} as
\begin{quote}
``privileged aspects that provide the classifier with `scales' to diversify systems along one or more dimensions.''
\end{quote}
The classic term to refer to such aspects is \emph{gestalt}, namely the
``essence or shape of an entity's complete form''~\cite{Jacks10}.
The accent is thus on a system's salient traits rather than on its architectural composition
or its organizational design.
%Through gestals, systems are classified
%according to several attributes---for instance  anatomical (architectural), physiological (organizational),
%or behavioral.

An important consequence of gestalt-based classification methods is the fact that they
function as \emph{models}:
they highlight certain aspects or features of a system while hiding others.
As an example, the behavioral gestalt introduced in~\cite{RWB43} only focuses on
\begin{quote}
``the examination of the
output of the [system] and of the relations of this
output to the input. By output is meant any
change produced in the surroundings by the
[system].  By input, conversely, is meant any
event external to the [system] that modifies this
[system] in any manner.''
\end{quote}
This results in generic classes that include very different systems---for instance  natural systems,
artificial, computer-based systems, and
business bodies\footnote{Boulding's class, for instance,
	include among others ``Clockworks'', ``Thermostats'', ``Cells'', ``Plants'', ``Animals'', and
	``Transcendental Systems'', which are generic names that may refer to systems of any nature.}.
From a practical point of view, systems in a class are considered as equivalent representatives
of their class---as it is the case in equivalence classes in algebra~\cite{Sprugnoli}. This is
exemplified by the relation of ``Boulding-equivalence'' introduced in~\cite{DF12a}.

\begin{definition}[Systemic classes]
Given any GST $T$ defining $n>1$ classes of systems
according to a given gestalt $g$, we shall call $T$-equivalence the equivalence relation
corresponding to the $n$ classes of systems. Those classes shall be called ``systemic classes according to $T$ and $g$'',
or, when this may be done without introducing ambiguity, simply as ``systemic classes.''
\end{definition}

Moreover, traditional system classifications pay little or no attention to the collective nature
of systems. In other words, systems are mostly considered
as individual, monolithic entities instead of the result of an organization of parts,
each of which is in itself another system.

As we have shown in~\cite{DF14b}, this translates in a \emph{partial order\/} among systems:
systems may be practically compared with one another---for instance, as of their intrinsic
resilience~\cite{DF15b}---only if they belong to different systemic classes. There is no easy
way to tell which of two Thermostats, or for instance two Cells\footnote{As already mentioned,
Thermostat and Cell are the names of two classes of the Boulding-equivalence relation.},
is better suited to manifest a given emerging property.

In what follows we propose to tackle this problem by considering two ``dimensions'':
\begin{itemize}
\item A ``horizontal'' dimension, regarding the system as an entity
resulting from the organization of a number of peer-level individual components.
\item A ``vertical'' dimension, regarding the system as a collective entity
resulting from the social organization (\emph{sensu}~\cite{Bou56}) and cooperation of a number of organs,
each of which is also socially organized into a collection of other organs~\cite{DBLP:journals/corr/Florio15c}.
\end{itemize}

%--------------------------------------------------------------------------%
\subsection{Horizontal dimension of system classification}\label{s:clas:hor}
%--------------------------------------------------------------------------%
Our starting point here is the conjecture that most of the classes introduced
in GST's may be described in terms of the five 
components of the so-called MAPE-K loop of autonomic
computing~\cite{KeCh:2003}, corresponding respectively to
\begin{description}
\item[\M:] the ability to perceive change;
\item[\A:] the ability to ascertain the consequences of change;
\item[\P:] the ability to plan a line of defense against threats deriving from change;
\item[\E:] the ability to enact the defense plan being conceived in step 3;
\item[\K:] the ability to treasure up past experience and continuously improve, to some extent, 
abilities \M--\E.
\end{description}

\begin{definition}[Systemic features]
As we have done in~\cite{DF14b}, in what follows we shall refer to abilities \M, \A, \P, \E, and \K{}
as to a system's \emph{systemic features}.
\end{definition}

As an example of the expressiveness of a system classification based
on the above systemic features, it is easy to realize that
the systemic class of purposeful, non-teleological systems~\cite{RWB43},
corresponding to Boulding's Thermostats, can also be interpreted as
the class of those systems that are
characterized by very limited perception (\M), analytical (\A), and operational (\E) quality
and by the absence of planning (\P) and learning (\K) ability.
Another example is given by the systemic class of extrapolatory systems, which
roughly corresponds to Boulding's class of Human-Beings. Systems in this class
possess complex and rich systemic features \M--\K.

As already mentioned, an intrinsic problem with systemic classes is that all of the systems
in a class are evened out and equalized. Obviously this is problematic, because systemically equivalent systems
may be in fact very different from each other. Two Thermostats may base their actions
on different context figures---think for instance of an accelerometer and a gyroscope
when used for fall identification~\cite{DBLP:journals/corr/FlorioP15b,EDC2013}.
Two Human-Being systems may have different analytical, planning, or learning features
due to, e.g., different design trade-offs\footnote{Explanations and examples
	of those trade-offs in natural systems can be found,
	e.g., in~\cite{Nilsson08,Nilsson08b,WeWi55,Nilsson14}.}.

Mapping the existing GST's onto the five systemic features allows
for a finer differentiation if we further decompose each class into sub-classes.
A way to do this has been described, for perception and analytical organs,
in~\cite{DF12a} and for planning organs in~\cite{DF13b}.

The idea is to either detail the quality of a systemic feature or to identify
the systemic class of the corresponding organs.

%In the cited papers we distinguished the case of perception from that 
For perception, the quality of \M{} is made explicit---to some extent---by specifying which
subset of context figures is perceived by \M. Notation ``$\M(M)$'' is then used
to state that perception is restricted to the context figures
specified in set $M$.
In next subsection we show how this makes it possible to use simple Venn diagrams to compare the perception feature
in systems and environments.

%verify whether any two
%systems have comparable qualities of \M and if so, which of the two systems has
%a wider perception~\cite{DF12a,DF13b}. In the cited reference we also showed how
%such method may reveal whether a system matches the requirements of a deployment
%environment and even how to compare  whether
\subsection{Perception}

\begin{figure*}[ht!]
\begin{center}
\subfigure[Perception of systems $a$ and $b$ with respect to that of hypothetical perfect system $m$
in the case of~\eqref{e:notsubset}.
The intersection region represents the context variables perceived by both $a$ and $b$.]{%
\label{f:perce}
\includegraphics[width=0.44\textwidth]{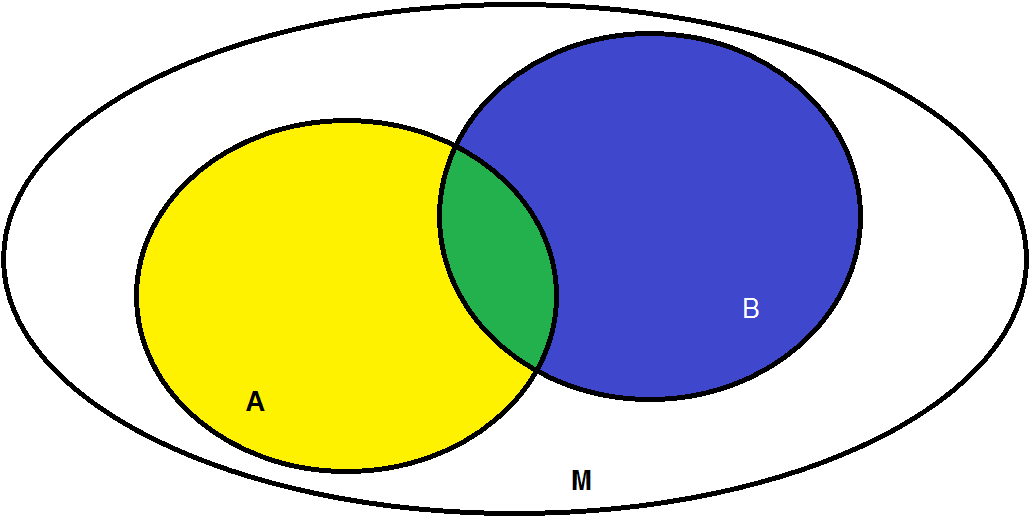}%
}
\hspace*{4pt}
\subfigure[Perception of $a$ and $b$ when~\eqref{e:subset} is valid.
In this case we can state that $a \PREC b \PREC m$: the perception feature of $a$ is
less than $b$'s,
which in turn is less than $m$'s.]{%
\label{f:perce2}
\includegraphics[width=0.44\textwidth]{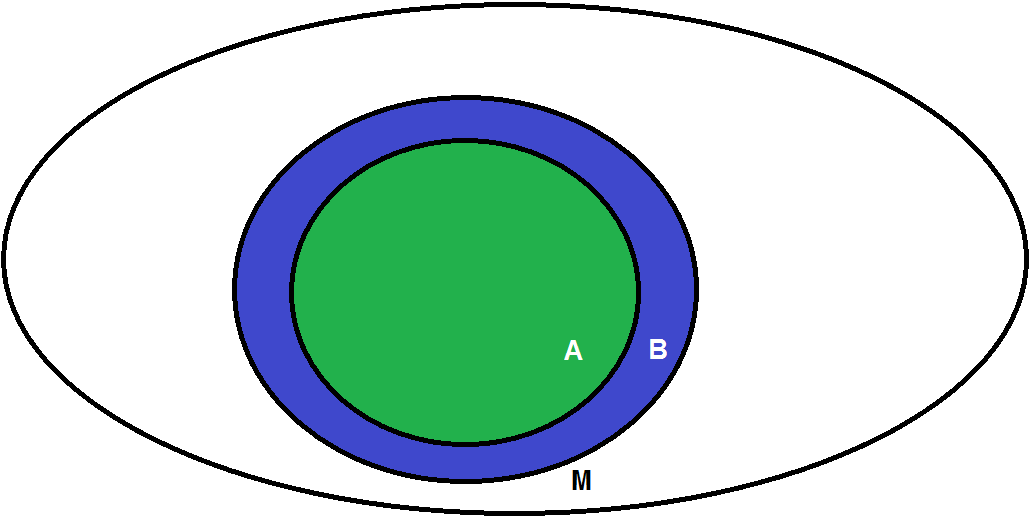}%
}%  ------- End of the first row ----------------------%
\end{center}
\caption{Venn diagrams are used to reason about the systemic features of perception in two systems.}
\end{figure*}

Let us consider any two systems $a$ and $b$, respectively characterized by $\M(A)$ and $\M(B)$.
There can be two cases: either 
\begin{equation}
(A\subset B) \lor (B\subset A) \label{e:subset}
\end{equation}
or otherwise, namely
\begin{equation}
(A\not\subset B) \land (B\not\subset A).\label{e:notsubset}
\end{equation}
As we showed in~\cite{DF13b}, if \eqref{e:subset} is true and in particular \( A \subseteq B \),
then we shall say that $b$ is endowed with a greater perception than $a$.
Notation
$a \PREC b$ will be used to express this property.
Likewise if \eqref{e:subset} and \( B \subseteq A \) then we shall have that $b \PREC a$.
This is exemplified in Fig.~\ref{f:perce2}, in which 
\begin{equation}
A\subseteq B\subseteq M, \label{e:monad}
\end{equation}
the latter being the set of all the possible context figures.
Clearly no system $m$ such that $\M(M)$ exists,
though we shall use of it in what follows as a reference point---a hypothetical
system endowed with ``perfect'' perception and
corresponding to the ``all-seeing eye''
of the \emph{monad}, which ``could see reflected in it all the rest of creation''~\cite{leibniz2006shorter}.

Expression~\eqref{e:monad} tells us that
$a$, $b$, and $m$ are endowed with larger
and larger sets of perception capabilities.
Expression \(a\PREC b\PREC m\) states such property.

%We deem it important to highlight how perception spectra such as set $A$ and $B$
%should be
%actually represented as functions of time; the mission characteristics; and the current context.
%In other words, perception should not be taken as an absolute and immutable feature but rather as
%the result of several dynamic processes, e.g., the current state of the sensory subsystem,
%the current quality of their services, as well as how the resulting times, throughputs, failures, and
%latencies match with the current mission requirements.
%For the sake of simplicity we shall nevertheless refer to perception spectra simply as sets.

A similar approach may be 
used to evaluate the environmental fit
of a given system with respect to a given deployment environment.
%---for instance, to gain insight in the match between that system and its intended execution environment.
As an example, Fig.~\ref{f:perce} may be also interpreted as a measure of the
perception of system $a$ with respect to the measure of perception
\emph{called for\/} by deployment environment $b$. The fact that $B\setminus A$ is non-empty
tells us that $a$ will not be sufficiently aware
%of the context changes occurring in $b$. Likewise $A\setminus B\neq\emptyset$ tells us that $a$ is designed
of the context changes occurring in $b$. Likewise $A\setminus B\neq\varnothing$ tells us that $a$ is designed
so as to be aware of figures that will \emph{not\/} be subjected to change while $a$ is in $b$.
The corresponding extra design complexity is (in this case)
a waste of resources in that it does not contribute to any improvement in resilience or survivability.

Finally, Venn perception diagrams may be used to compare environments with one another. This may be
useful especially in ambient intelligence scenarios in which some control may be exercised on
the properties of the deployment environment(s).

Estimating shortcoming or excess in a system's perception capabilities provides useful
information to the ``upper functions'' responsible for driving the evolution of that system. Such functions
may then make use of said information to perform design trade-offs among the resilience layers.
As an example,
%(see more on this in Sect.~\ref{s:rhm}). As an example,
a system able to do so may reduce its perception spectrum and use the resulting complexity budget to widen both
its \A{} and \P{} systemic features.

\subsubsection{Limitations of our approach}
Although effective as a secondary classification system, our
approach is also a model---in other words,
a simplification. In particular, reasoning merely in terms of subsets of context figures
underlies the unlikely assumption of an at-all-times perfect and at-all-times reliable
perception organ.
Furthermore, our approach does not take
into account the influence that other organs may have on the perception organ\footnote{As
	illustrated in, e.g.,~\cite{BrGr15}, 
	perception may be misled by higher functions; as an example, the analytical organ
	may provide an interpretation of the ongoing facts that may lead the perception
	organ into ``concealing'' certain facts or overexposing others.}.

\subsection{Other systemic features}\label{s:osf}
The above approach based on Venn diagrams cannot be applied to systemic features \A--\K.
An alternative approach was suggested in~\cite{DF13b}. The idea is to select
a GST $T$ and
``label'' each organ with its systemic class in $T$.
This allows a finer classification to be obtained and a greater differentiation
of systems in the same systemic class.

An exemplary way to apply this method
is shown in~\cite{DF14b} by making use of the classic behavioral method. Thus for instance
the organ responsible for planning responses to context changes---corresponding thus
to systemic feature \P---may be characterized as belonging to, e.g., the ``Thermostat'' systemic class.
As an example, our adaptively-redundant data structures~\cite{DF13a} belong to
the systemic class of predictive mechanisms, although their \P{} organ belongs
to the simpler class of purposeful, non-teleological systems.

In certain cases, instead of
a GST, one could use an existing classification peculiar of a given systemic feature.
Lycan, for instance, suggests
the existence of at least eight apperception classes~\cite{Lycan96} (namely, eight
\A{} classes).

%------------------------------------------------------------------------%
\subsection{Vertical dimension of system classification}\label{s:clas:ver}
%------------------------------------------------------------------------%
\epigraph{%
``The Internet is a system --- \\and any system is an internet.''}
{\url{https://goo.gl/WTnvLD}}

As we already mentioned, a classic assumption shared by several GST's
is that of describing systems from an individual perspective. Our ``horizontal''
classification proposes a first solution to this deficiency by providing a top level view
to a system's organization. By exposing the main organs \M--\K{} we provide a more
detailed information about the nature and features of the system at hand.

Our vertical classification goes one step further.
It does so by regarding systems as collective entities
resulting from the social organization (\emph{sensu}~\cite{Bou56}) and cooperation of a number of organs,
each of which is also socially organized into a collection of other organs.
As in Sect.~\ref{s:clas:hor} systems were exposed as systems-of-systems, similarly here
we model systems as network-of-networks. Better, systems are interpreted here as networks of nodes,
each of which is in itself another network of nodes. As discussed in Actor-network Theory~\cite{Latour06},
each node ``blackboxes'' and ``individualizes'' its network by assuming the double
identity of individual and collective system---a concept that finds its sources in the
philosophy of Leibniz~\cite{DF14c}.

Being a system, 
each node is eligible to belong to a systemic class. The horizontal classification
introduced in Sect.~\ref{s:clas:hor} may therefore be applied:
a given node may for instance
behave as an object~\cite{RWB43} and thus be perception-free; or it may be a Thermostat or
a ``Servomechanism'' (thus with limited perception and no analytic functions); or it
may be an organ, as it is the case in Boulding's Cells. In such a case, it
may be endowed with perception and limited analytical capabilities. Moreover,
it may be an organism (a Plant or an Animal) and be endowed with extended perception,
some analytical capabilities, and limited planning capabilities. At the top of the scale,
it may be a self-conscious system (Boulding's Human-Beings) and rank high on all
the systemic features.

\begin{definition}[Systemic level]
Given any system vertically classified into a network of nodes, we shall refer to each set of nodes
that are peer levels as to a \emph{systemic level}.
\end{definition}
An example of systemic level is given by the top level view resulting from our
horizontal classification. Another example is
	the very root of the vertical classification, namely the individual, ``holistic''
representation of the system---although of course in this case the systemic level is a singleton.

%----------------------------------%
\subsection{Preliminary conclusions}
%----------------------------------%
In this section we have introduced a horizontal and a vertical system classification as a tool to further
differentiate systems belonging to a same GST class.
By making use of our proposed classifications any system 
is organized both vertically and horizontally: vertically, as a network of nodes; and horizontally,
as an organization of peer-level organs corresponding to the system's five systemic features.

We deem important to highlight how, by means of our classifications,
systems expose their structure of complex networks of systems-within-systems,
or equivalently of network-of-networks. This translates into 
a Matryoshka-like structure corresponding to the class of networks
known as \textbf{nested compositional hierarchies} (NCH).

NCH have been defined in~\cite{HT:TE14a} as ``a pattern of relationship among entities based on the
principle of increasing inclusiveness, so that entities at one level are composed of parts
at lower levels and are themselves nested within more extensive entities''.
The class of NCH organizations
is widespread in natural systems because of its straightforward support of modularity---in
turn, an effective way to deal with complexity and steer evolvability~\cite{WaAl1996}.
Further discussion on this may be found, e.g., in~\cite{DF15c}.

Finally, we remark how vertical organization and NCH produce a
\emph{fractal organization\/} of parts in a variety of levels, or scales. In
natural systems those scales range from the microscopic, sub-atomic 
to the macroscopic level as typical of, e.g., biological ecosystems.
When classifying systems in order to compare their systemic characteristics a trade-off
shall be necessary
in order to limit the vertical expansion to a practically manageable number of levels.

More information and an example of fractal organization are provided in Sect.~\ref{s:LS}
and Sect.~\ref{s:fso:fso}.

%%%%%%%%%%%%%%%%%%%%%%%%%%%%%%%%%%%%%%%%%%%%%%%%%%%%%%%%%%%%%%%%%%%%%%%%%%%%%%%%%%%%%%%%%%%%%%%
\section{Making use of our classification system to assess and compare resilience}\label{s:res}
%%%%%%%%%%%%%%%%%%%%%%%%%%%%%%%%%%%%%%%%%%%%%%%%%%%%%%%%%%%%%%%%%%%%%%%%%%%%%%%%%%%%%%%%%%%%%%%

Let us consider two examples:
\begin{enumerate}
\item A bullet passing through the body of a living being.
Such a traumatic event directly affects a number of organs and systems of that being.
Interdependence among organs and systems is likely to lead to cascading effects
that may in turn lead to severe injuries or the loss of life.
\item As a second example, let us consider the case of a hurricane hitting a region.
Catastrophic events such as this one typically ripple across the network-of-nodes
triggering the concurrent reactions of multiple crisis management organizations~\cite{CARRI3,RAND}.
\end{enumerate}

The above two cases exemplify what we conjecture may be a ``general systems law'':
any catastrophic event that manifests itself within a system's boundaries
creates a critical condition that \emph{crosscuts\/} all of that system's
levels and nodes, with consequences that can affect the nodes that are directly hit
as well as those depending on them. Consequences may ripple through the boundaries
of the system and lead to chains of interconnected local failures possibly bringing
to general system failure.

In fact, catastrophic events such as the two exemplified ones
reveal a system's true nature and organization---as litmus paper
does by unravelling the pH value of a chemical solution~\cite{Litmus}.
The illusion of an ``in-dividual'' (non-divisible) system is shuttered and replaced by the awareness
of the fragmented nature of the system as a system-of-systems and a network-of-networks.

The adoption of a horizontal and a vertical system classification offers in this case
a clear advantage, in that it provides a view to the actions that may be
expected from each of the involved constituent systems.
Depending on each system's systemic level, the reaction to the catastrophic event might include
different flavors of
perception steps; analytical steps; planning steps; reaction
plan execution steps; and knowledge management steps (namely, knowledge feedback and its persistence).
Conditional ``might'' is used here
because, as mentioned already, not all the involved systems may have a complete set of systemic features
and the corresponding organs may have different systemic classes.
Thus for instance the catastrophic event and its ripples will only be perceived by systems
whose perception organs include the
context figures related to that event. As another example, a \P{} organ may produce a
response plan ranging from predefined responses up to reactive, extrapolatory,
and even antifragile behaviors~\cite{DF15b}.

Other factors may play a key role in local and overall responses to catastrophic events.
Those responses may depend, e.g., on the quality and performance of the involved organs. Said quality 
may be modeled as a dynamic system and expressed in terms of fidelity and its fluctuations
(called ``driftings'' in~\cite{DBLP:journals/corr/FlorioP15,DF14a}).
Moreover, as responses call for energy and energy being often a limited and valuable commodity, responses
enacted by some nodes are likely to exact energy from other nodes\footnote{For
	example, ``inner'' systems' action may deprive outer systems of their resources;
	and likewise outer systems' decisions may lead to poor 
	choices affecting the resources and the operational conditions of inner nodes.}.

Other important factors in the emergence of resilience
are given by what we commonly refer to as ``experience'' and ``wisdom'', which
correspond to systemic feature \K. Those factors are in some cases of key importance,
as they may lead to situations in which two identical systems reach very different degrees
of resilience~\cite{Washburn15}.

A final and very important aspect that is not considered by our classification system is given
by harmony and cohesion between the ``parts'' and the ``whole''. This is exemplified by the famous
apologous that Menenius Agrippa gave the commons of Rome during the so-called ``Conflict of the Orders''~\cite{Lily}.
In his speech Agrippa imagines a disharmony among the parts of the human body, with ``busy bee'' organs
complaining about the less active role played by other organs. Because of the discord,
the more active parts undertake a strike, whose net result is a general failure because in fact
all parts are necessary and
concur to the common welfare according to their role and possibility\footnote{See
	for instance~\cite{leibniz2006shorter}:
	``There
	is always in things a principle of determination which
	must be sought in maximum and minimum; namely, that
	the greatest effect should be produced with the least
	expenditure, so to speak.''}.
In other words, disharmony is a disruptive
force that breaks down the whole into its constituent parts. Resilience may very well be affected
in the process, as exemplified by a nation unable to effectively respond to an attack because of the lack
of identification of its citizens with the state.

%--------------------------------------------------%
\subsection{Resilience as an interplay of opponents}
%--------------------------------------------------%
In our previous work~\cite{DF15b} we discussed resilience as
the emerging result of a dynamic process that represents the dynamic interplay between the behaviors
exercised by a system and those of the environment it is
set to operate in. With the terminology introduced in this paper we may say
that resilience is the result of the effects of an external event on a system's
horizontal and vertical organization. The external event manifests itself at all
systems and networks levels and activates a response that is both individual and
social. As we conjectured in the cited reference, game theory (GT)~\cite{EaKl10} may provide a convenient
conceptual framework to reason about the dynamics of said response.
GT players in this case are represented by nodes, while GT strategies represent the plans
devised by the nodes' \P{} organs. As suggested in our previous work,
a way to represent the strategic choices available to the GT players
is to classify them as behaviors.
As an example, if node $n$ is able to
exercise extrapolatory behaviors, then $n$ may in theory choose between the following strategies
of increasing complexity:
random; purposeful/non-teleological; teleological/non-extrapolatory; or extrapolatory~\cite{RWB43}.
In practice, the choice of the strategy shall also be influenced by some ``energy budget''
representing the total amount of consumable resources available system-wide to enact the behaviors of
all nodes. Said energy budget would then serve as a global constraint
shared by all of the nodes of the system across both the horizontal and vertical organizations.

GT payoffs could then be associated to the possible exercised behaviors, with costs (in terms
of consumed energy budget resources) proportional to the complexity of the chosen behavior.

It seems reasonable to foresee that the adoption of GT as a framework for discussing the
resilience of systems classified according to our approach
shall require the definition of nested compositional hierarchies of payoff matrices---sort
of interconnected and mutually influencing payoff ``spreadsheets''.

%%%%%%%%%%%%%%%%%%%%%%%%%%%%%%%%%%%%%%%%%%%%%%%%%%%%%%%%%%%%%%%%%%%%%%%%%%%%%%%%%%%%
\section{An intelligent environment based on our system classification}\label{s:fso}
%%%%%%%%%%%%%%%%%%%%%%%%%%%%%%%%%%%%%%%%%%%%%%%%%%%%%%%%%%%%%%%%%%%%%%%%%%%%%%%%%%%%
In Sect.~\ref{s:res} we have considered resilience, interpreted as the outcome of
a \emph{conflict\/} between two opponents. We have shown that our system classification
allows said conflict to be detailed within the systems boundaries along
their vertical and horizontal dimensions.

A dual consideration can be made by considering other emerging properties---for instance,
performance, safety, and quality of experience. An intelligent environment may be designed with the aim to 
assist a system to achieve its design goals by structuring it \emph{after\/} the horizontal and vertical
classification of that system. One such system is the middleware designed in the framework
of project ``Little Sister''. In what follows
we briefly introduce some elements of that project that are useful to the present discussion
and then
we suggest how the architecture of the LS middleware may facilitate the expression of
optimal services combining emerging properties such as the above mentioned ones.

%------------------------%
\subsection{Little Sister}\label{s:LS}
%------------------------%
Little Sister (LS) is the name of a Flemish ICON project financed by
the iMinds research institute and the Flemish Government Agency for Innovation by
Science and Technology. The project run in 2013 and 2014 and aimed to deliver a
low-cost telemonitoring~\cite{Meystre05} solution for home care.
Cost-effectiveness was sought by replacing expensive and energy-greedy smart cameras
with low-resolution cameras
based on battery-powered mouse sensors~\cite{IGO}.

%The low resolution of the LS sensors inherently guaranteed---to some extent---privacy, which brought to the name
%of the project as a

\begin{figure*}
\centerline{\includegraphics[width=1.0\textwidth]{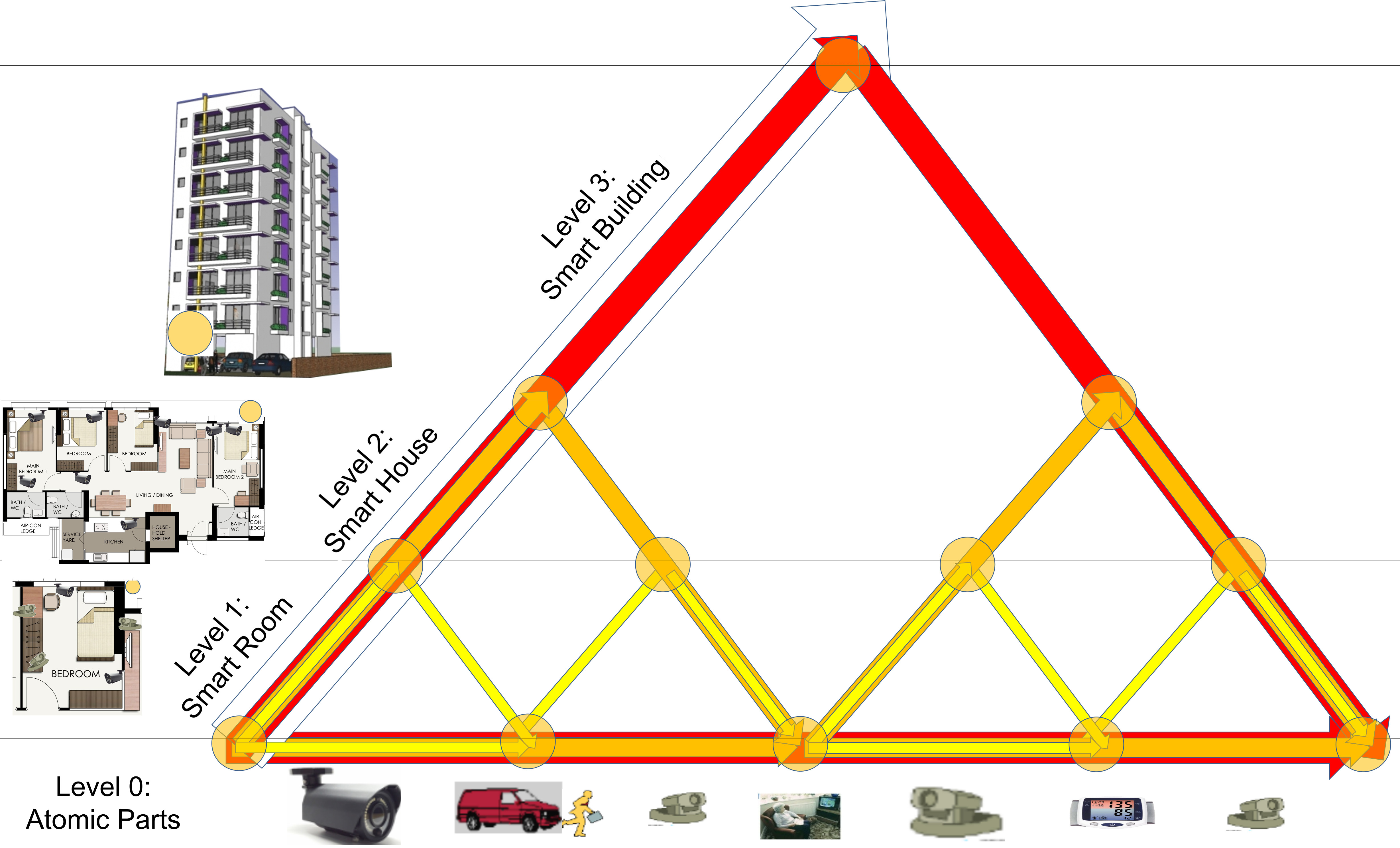}}
\caption{Exemplification of the LS software architecture.}
\label{f:lsfso}
\end{figure*}

The LS software architecture is exemplified in Fig.~\ref{f:lsfso}. As suggested by the shape
of the picture, LS adopts a fractal organization
in which a same building block---a web services middleware component---is repeated across
the scales of the system. In fact the vertical classification of the LS service is, in a sense,
\emph{revealed\/} through the fractal organization~\cite{Koe67,Warnecke93,Tharu96,TWN1998}
of the LS software architecture:
\begin{itemize}
\item Atomic constituents are grouped into a ``level 0'' of the system. Those constituents are
wrapped and exposed as manageable web services that represent a periphery of \M{} and \E{} nodes.
\item Said web services constitute
a first level of organs that manage the individual rooms of a smart house under the control
of a middleware component responsible for systemic features \A{} and \P.
\item Individual rooms are also wrapped and exposed in a second level under the control of
the same middleware component, here managing a whole smart house.
\item The scheme is repeated a last time in order to expose smart house services, also
under the control of our middleware component. This third level is called smart building level.
\end{itemize}

No systemic feature \K{} is foreseen in LS.

As evident from the above description, the LS system represents a practical
example of our horizontal and vertical classification. The levels of the LS
software architecture allow services to be decomposed into
\begin{itemize}
\item Low-level services for context change identification;
\item Medium-level services for situation identification~\cite{YeDM12};
\item High-level services for overall system management,
\end{itemize}
which naturally leads to the choice of a three-level vertical classification.

%-------------------------------------------------------------------------%
\subsection{A fractally organized intelligent environment}\label{s:fso:fso}
%-------------------------------------------------------------------------%
As mentioned at the beginning of this section, awareness of a system's horizontal and vertical
classification may be exploited to create an environment reflecting the structure of that system
and designed in order to provide assistive services to that system. In what follows we provide
an example of such an environment, implemented through the LS middleware.

As the system is structured into three levels so also our middleware is organized on three levels.
A same middleware module resides 
in the three environments that represent and host the nodes of each level: rooms, houses, and building,
and corresponding respectively to levels 1--3 in Fig.~\ref{f:lsfso}. The middleware
wraps sensors and exposes them as manageable web services. These services are
then structured within a hierarchical federation~\cite{Oasis1}. 
More specifically, the system maintains
dedicated, manageable service groups for each room in the building,
each of which contains references to the web service endpoint of the
underlying sensors (as depicted in level 0 and 1 in Fig.~\ref{f:lsfso}).  These
``room groups'' are then aggregated into service groups
representative of individual housing units. Finally, at the highest
level of the federation, all units pertaining to a specific building
are again exposed as a single resource (level 3).  All services and
devices situated at levels 0--3 are placed within the
deployment building and its housing units; all services are exposed as
manageable web services and allow for remote reconfiguration.

By exposing the sensors as manageable web services and by means
of a standardized, asynchronous publish-and-subscribe mechanism~\cite{Oasis2} the
middleware ``hooks'' onto the system's perception and executive organs---namely
those organs corresponding to systemic features \M{} and \E.
All status and control communication is thus transparently received by the middleware, which 
checks whether the received information calls for functional adjustments or if it
represents a safety-critical situation requiring a proper response.

Each response is managed by the middleware
through a protocol that requires the cooperation of ``agents'' (system nodes).
As in data-flow systems~\cite{Sharp:1992:DFC:140762,Tom67}
it is the presence of the input data that ``fires'' an operation, likewise in LS protocols
it is the presence of all the required roles
that enables the launch of a protocol. For this reason we refer to our
approach as to a \emph{role-flow scheme}.

Said role-flow scheme of the LS middleware is a simplified version of the
more general strategy introduced in~\cite{DF13c,DeBl10},
in which nodes publish a semantic description of the roles they may play and the services they
may offer. Semantic matching is then used in the enrollment phase~\cite{SDB13a}.

The above sketched distributed organization, in which a same building block is repeated
in a nested compositional hierarchy of nodes,
is known in the literature as a \emph{fractal organization}~\cite{Koe67,Warnecke93,Tharu96,TWN1998}.
``Canon'' is the term used to refer to a fractal organization's building block.
Each node of the hierarchy hosts a canon---which in the case at hand is implemented by our middleware module.

It is important to highlight how the canon at level $i$ is both a node of that level and
a node of level $i+1$ (if $i$ is not the top level). As a node of level $i$, the canon
plays the role of that level's ``controller'' by executing the role-flow scheme.
At the same time, canon $i$ represents and
``punctualizes''~\cite{Law92} the whole level $i$ into a single level $i+1$ node.

A peculiarity of the fractal organization of our middleware
is the interoperability and cooperation between its levels---a feature that is achieved through the concept of 
\emph{role exception}. When middleware module at level $i$ does not find all the roles
required to launch a protocol, it declares
an exception: being also a node of level $i+1$, its status and notifications are transparently
published and received by the middleware module at level $i+1$. The latter thus becomes
aware of a level $i$ protocol that is missing roles.
Missing roles are thus also sought into the parent node, and from there into the parent's parent node,
and so on.

A consequence of this strategy is that roles are first sought in the level where a ``need'' has
arisen; only when that level fails to answer the need, the hierarchy is searched in order
to complete the enrollment and launch the protocol. The result of this strategy is a new,
trans-hierarchical ``temporary organ'', consisting of the nodes in any level of the hierarchy
that best-match the need at the time of enrollment.

Since the new organ includes nodes from multiple levels of the network-of-networks, we call the new
organ a
social overlay network (SON). Fractal social organization (FSO) is the name we gave to
a fractal organization implementing the above strategies~\cite{DF13c}.

%.........................................................................%
\subsubsection{Adaptive dimensioning of response protocols}\label{sss:dtof}
%.........................................................................%
The same algorithm employed for the adaptively-redundant 
data structures mentioned in Sect.~\ref{s:osf}
was adopted for the LS middleware. In what follows we briefly describe that algorithm.

As mentioned above, the middleware becomes timely aware of the state of the LS system.
This includes the definition of the current ``situation''.
Situations~\cite{YeDM12} range from low-level
context changes pertaining to the state of devices (for instance, a sensor's battery level reaching a
given lower threshold) up to high-level, human-oriented conditions and events.
An example of the latter case is
situation $s_1=$ ``the resident has left her bed during the night and is moving towards the kitchen''.
Another example is $s_2=$ ``the resident is sleeping in her bed''.
In general, different situations call for different reaction protocols, in turn calling for a different
amount of nodes and resources. For instance, it may be sensible to appoint more resources
to situation $s_1$ than to $s_2$.
Our algorithm implements
%The rationale of this algorithm is that different situations require different
%amounts of resources in order to guarantee the system's design goals.
%
%This resulted in
an adaptive dimensioning strategy that estimates the amount of
nodes best-matching the current situation with minimal impacting on the system's design goals.

In the absence of activity and when the current situations are assessed as relatively
stable and safe---as it appears to be the case in situation $s_2$---the
middleware gradually
decreases its requirements down to some minimum threshold. This threshold level is
estimated beforehand so as to still guarantee prompt reaction
as soon as variations are detected in the ongoing scenarios.

In a sense, the LS middleware
tracks the activity of the residents closely \emph{imitating\/}  their behaviors: when a resident,
e.g., sleeps, the
corresponding LS entity also goes-to-sleep (or better, it goes to low consumption mode).
On the contrary, when the residents awake or are in need, the
LS entity also goes to full operational mode.

As already mentioned,
the gradual adjustments of the LS operational mode is based on an algorithm of autonomic redundant replicas selection.
At regular time steps the middleware component responsible for the current level
checks whether the current allocation was overabundant or underabundant
with respect to the ongoing situation. In the former case---namely if ``too many'' resources were employed,
the container selects some of the enrolled nodes and ``frees'' them. In the latter case,
either a ``better'' selection
of the same amount of fractals is attempted, or new fractals are enrolled, or both, by
following the strategy depicted in~\cite{BDB11a,BDFB12}.
A ``distance-to-failure'' function
is computed at each step to measure how the current configuration matched the current situation. The value of this
function determines overabundance vs. underabundance and the corresponding decrease vs. increase of the employed resources.

\begin{figure*}
\centerline{\includegraphics[width=0.7\textwidth]{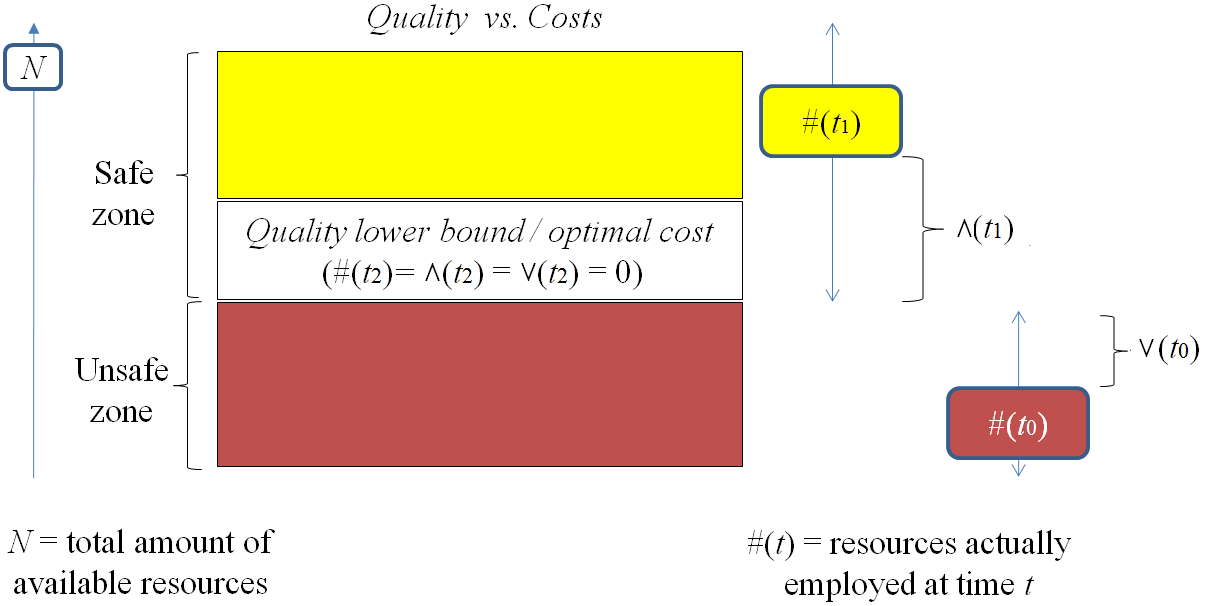}}
\caption{LS optimization is based on measuring resource overshooting (function $\wedge$) and
undershooting (function $\vee$) and adjusting resource allocation accordingly. The picture
shows the three possible system states: in $t_0$ the system experiences resource undershooting; 
in $t_1$, resource overshooting; and in $t_2$ it reaches an optimal resource expenditure.}
\label{f:opt}
\end{figure*}

The logic of this algorithm is graphically represented in Fig.~\ref{f:opt}. In such picture,
$N$ is the total amount of nodes
available (e.g., 10 sensors deployed in different positions in a resident's bedroom)
and $\hbox{\#}(t)$ is the amount of
``fired'' (namely, activated) sensors. If we assume that the current
situation, say $s$, will not change during a certain observation interval $T$ (because, for instance, the resident is
sleeping in her bed), then during $T$ we will have two stable ``zones'' corresponding to the different
allocation choices enacted by LS.
\begin{itemize}
	\item The unsafe zone is depicted as a red rectangle and represents choices corresponding
		to resource undershooting:
		here too few nodes were allocated with respect to the situation at hand.
		For any $t\in T$,
		function $\vee(t_0)$ tells us how big our mistake was at time $t_0$: how far we were at $t_0$ 
		from the minimal quality called for by $s$.
	\item The safe zone is given by the the white and the yellow rectangles.
		\begin{itemize}
		\item In the yellow rectangle
		the allocation was overabundant: too many resources were allocated (resource overshooting).
		For any $t_1\in T$ function $\wedge(t_1)$ tells us how large our overshooting was at time $t_1$.
		It also represents how far we were from the unsafe zone.
		\item The white rectangle represents the best choice: no overshooting or undershooting
			is experienced, which means that
			the allocation matches perfectly situation $s$.
			Here $\vee(t_2)=\wedge(t_2)=0$.
	\end{itemize}
\end{itemize}

The above mentioned ``distance-to-failure'' is then defined, for any $t$, as 
\begin{equation}
 \hbox{DTOF}(t) = \frac{\vee(t)}{N}.
\end{equation}
The allocation strategy of LS is based on tracking the past values of DTOF in order to estimate the ``best''
allocation of resources for next step. Regrettably, no implementation of the above design was completed
in the course of project LS, although a study of the performance of our strategy
is ongoing~\cite{Buys15}, with preliminary results available in the above cited papers.
%More information on this algorithm may be found in~\cite{BDFB12,DF13a,DB07a}.

\begin{figure*}
\centerline{\includegraphics[width=0.7\textwidth]{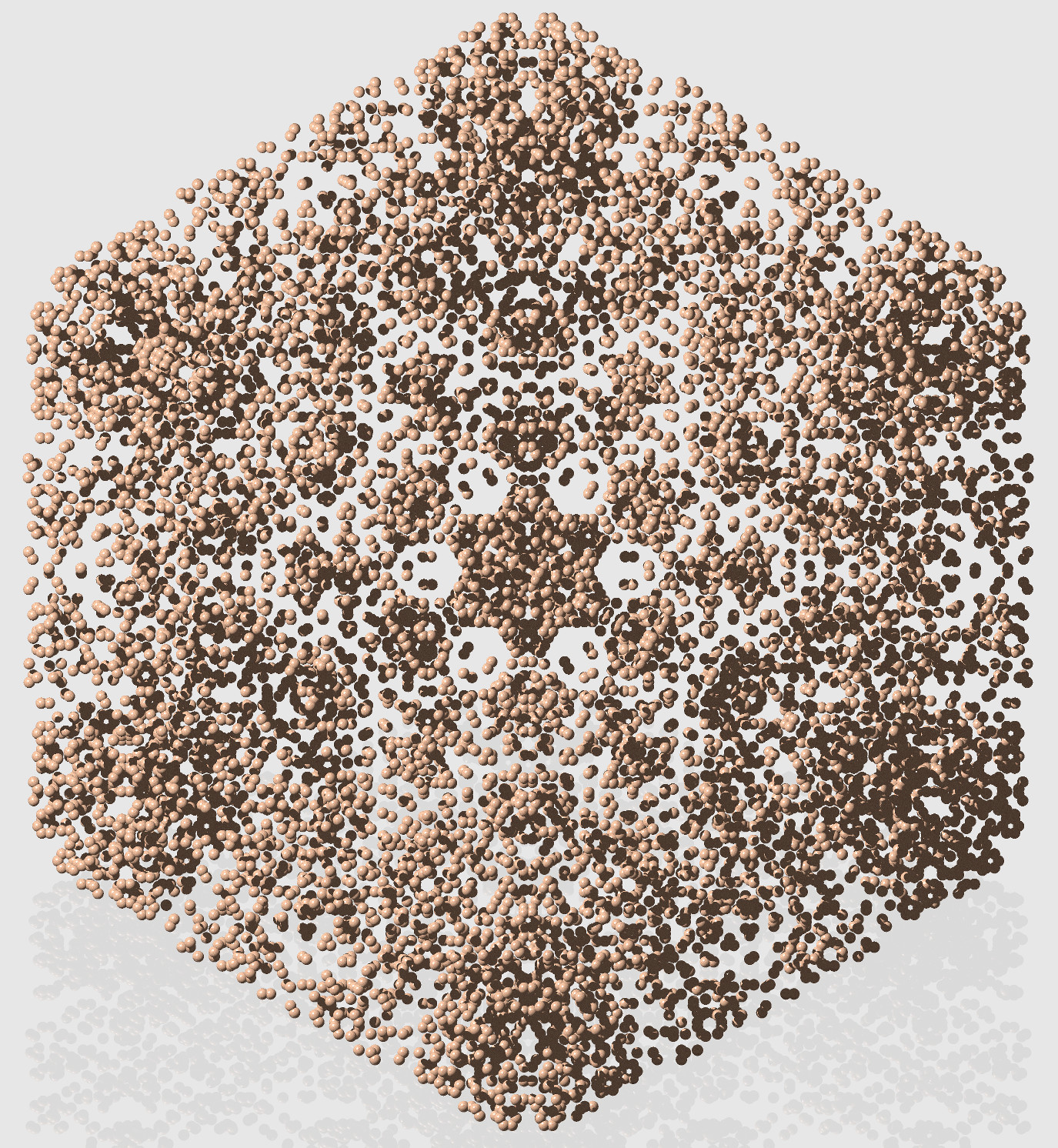}}
\caption{The graph represents the set of all the
social overlay networks that may emerge from an FSO with 9 nodes and 5 roles (roles 0--4).
   In this case the FSO consists of three nodes able to play role 0; three nodes able to play respectively roles 1--3;
   and three roles able to play role 4.
   The graph was produced with the POV-Ray ray tracing program~\cite{povray}.}
\label{f:sons}
\end{figure*}

Figure~\ref{f:sons} shows a three-dimensional representation of the space of all possible SON's that can
originate from an exemplary FSO.

%
%The LS middleware}
%Intelligent ambient whose goals are: safety; performance; energy-efficience.
%
%Not all of them reached, because of budget cuts and revised sub-goals...
%
%How? Through a NCH
%
%we designed a Fractal Social Organization~\cite{DF13c,DFSB13a}!
%Horizontal and vertical organization based on a building block
%repeated at all systems scales---the so-called Service-oriented Community~\cite{DeBl10}.
%
%
%QUI
%We define an intelligent environment as one that empowers / enlightens all its participants so that they may become effective components of a ``solution''. 
%
%
%
%
%An intelligent ambient may function as an organizational ``glue'' between the Parts and the Whole. We envision a fractal organization for such a intelligent ambient, and one that would ``mimic'' a system's network-of-network / system-of-systems structure (cf. LittleSister). (At least, the NoN structure)
%
%
%
%

%-------------------------------------------------------------------------%
\subsection{Environments as crosscutting optimizers and antifragility enablers}
%-------------------------------------------------------------------------%
We now briefly discuss the approach exemplified in Sect.~\ref{s:fso:fso} by considering
environments as crosscutting optimizers as well as enablers of antifragile behaviors.

%.................................................%
\subsubsection{Environments as crosscutting optimizers}
%.................................................%
We conjecture that environments such as the one
we have just sketched may function as \emph{crosscutting optimizers}: assisting environments
able to rapidly communicate awareness and wisdom from one level to the other of an
assisted system.
This is made possible by means of the mechanisms implemented by our FSO: exception, role-flow, and SON.
In FSO, unresolved local events are transferred automatically to the higher levels of the organization.
Local decisions and reactions are then exposed to the higher levels, and vice-versa: actions and decisions
occurring in the higher levels of the system may thus be perceived and analyzed by the ``inner systems'',
allowing those systems to understand the local consequences of ``global'' actions.

We conjecture that this may result in perception failures avoidance, reduced reaction
latency~\cite{Miskel08,Adair02}, increased agility, and avoidance of single-points-of-congestion.
Furthermore, the FSO enrollment does not discriminate between institutional and non-institutional nodes.
This encourages participation and collaboration and avoids community resilience failures
such as the ones experienced in the recovery phase after the Katrina and
Andrew Hurricanes.
The same non-discriminative nature makes it possible for
unnatural distinctions between, e.g., primary, secondary, and tertiary users, to be avoided~\cite{SDGB07a}.

%................................................%
\subsubsection{Environments as antifragility enablers}
%................................................%
As a second conjecture, we believe that environments based on our FSO may also function
as \emph{antifragility enablers}\footnote{%
	Antifragility is the term
	introduced by N. N. Taleb in~\cite{Taleb12} to refer to systems that are
	able to \emph{systematically\/}
	``enhance the level of congruence or fit between themselves
	and their surroundings''~\cite{stokols2013enhancing}. Quoting from Professor Taleb's book,
	\begin{quote}
	``Antifragility is beyond resilience or robustness. The resilient resists shocks
	and stays the same; the antifragile gets better.''
	\end{quote}
	An analysis of elastic, resilient, and antifragile behaviors was proposed in~\cite{DF15b}.}.

As we have already remarked, the LS middleware does not provide a complete implementation of
FSO. In particular, it does not foresee any component responsible for
the \K{} systemic feature. A major consequence of this is that the FSO enrollment in LS
is \emph{memoryless}: protocols are started from scratch, taking no account of their past ``history''.
Aspects such as the performance of a node as ``role player'' in the execution of a protocol; the
trustworthiness manifested by that node; the recurring manifestation of a same SON;
as well as its performance as executive engine for a protocol; were not considered in the LS design.
Of course nothing prevents to design a FSO in which the above and additional aspects are duly considered.
As an example, enrollment scores, telling which nodes best played a role in a given SON, may be
implemented by making use of algorithms of gradual rewarding and penalization such as the ones
described in~\cite{BDB11a,BDFB12,Buys15}. Those very same algorithms, applied at a different \emph{level},
may be used to gain wisdom as to the best-matching solutions. Proactive deployment of the best-scoring
SON's across the scales of the FSO may enhance its effectiveness in dealing with, e.g., disaster
recovery situations. Moreover, the resurfacing of the same transient SON's may lead to
\emph{permanentification}, namely system reconfigurations in which new permanent nodes and levels manifest
themselves. In other words, by means of the above and other antifragile strategies the system
and its vertical organization may \emph{evolve\/} rather than \emph{adapt\/} to the conditions expressed by a
mutating environment.

%Tutto ci\`o pu\`o far evolvere la struttura gerarchica del sistema; mutarla in dipendenza di situations
%endogene (condizioni ambientali) ed esogene (condizione dei residenti, delle risorse del sistema; 
%``performance'' dei vari nodi ed organi del sistema, etc)
%cos\`\i{} da raggiungere una sorta di antifragilit\`a di sistema e di ambiente.

%%%%%%%%%%%%%%%%%%%%%%%%%%%%%%%%%%%%%%%%%%%%%%
\section{Conclusions}\label{s:end}
%%%%%%%%%%%%%%%%%%%%%%%%%%%%%%%%%%%%%%%%%%%%%%
In the present work
we have proposed to augment existing GSTs by making use of a horizontal and a vertical dimension.
This introduces systemic subclasses that make it possible to further differentiate
systems belonging to the same GST class. We have shown how this allows for a finer comparison of
systems with respect to their ability to achieve their intended design goals.
In particular we have shown how to make use of our classification approach to
assess the resilience exhibited by a system when deployed in a target environment.
Building on top of our previous work on resilient behaviors, here we have further
discussed resilience as a property emerging from an interplay of the behaviors exercised
by two opponents. 

As a dual argument, here we have also considered properties emerging from
interplays of ``opposite sign''---namely, interplays between a system and an \emph{assisting\/}
(rather than an opposing) environment. We have discussed how our classification approach allows
for the creation of an environment mimicking a system's horizontal and vertical structure.
By doing so, the assisting environment realizes a ``systemic exoskeleton'' of sorts,
which is able to interface with that system's organs and artificially augment its
analytical, planning, and knowledge systemic features. In particular we have shown
how FSO and its concepts of exception, role-flow, and SON, realize inter-organizational
collaboration between nodes residing in any of the levels of the system organization.

Future work will include the simulation of scenarios
in which the environment plays either the role of opponent or that
of assistant, as  
discussed in this paper.
Preliminary results have been already obtained by simulating ambient assistive
environments~\cite{DeFPa15a,DBLP:journals/corr/FlorioP15b}. A theoretical discussion of resilience in the
framework of Game Theory is also among our plans.

\subsubsection*{Acknowledgments.}
This work was partially supported by
iMinds---Interdisci\-pli\-nary institute for Technology, a research institute
funded by the Flemish Government---as well as by the Flemish Government Agency for Innovation by
Science and Technology (IWT).
The iMinds Little Sister project was a project co-funded by iMinds with project support of IWT
(Interdisciplinary institute for Technology)
%a research institute founded by the Flemish Government.
Partners
%Companies
%and organizations
involved in the project are
Universiteit Antwerpen,
Vrije Universiteit Brussel,
Universiteit Gent,
Xetal,
Niko Projects,
JF Oceans BVBA,
SBD NV,
and Christelijke Mutualiteit vzw.

\bibliographystyle{spmpsci}

\end{document}